\pdfoutput=1
\documentclass[prd,twocolumn,eqsecnum,nofootinbib,floatfix]{revtex4}
\usepackage{amsmath}
\usepackage{bm}
\newcommand{\E}{{\cal E}} 
\newcommand{\B}{{\cal B}} 
\newcommand{\stf}[1]{{\langle #1 \rangle}}
\begin{document}
\title{Tidal interaction of black holes and Newtonian viscous bodies}  
\author{Eric Poisson} 
\affiliation{Department of Physics, University of Guelph, Guelph,
  Ontario, N1G 2W1, Canada;} 
\affiliation{Canadian Institute for Theoretical Astrophysics,
  University of Toronto, Toronto, Ontario, M5S 3H8, Canada} 
\date{July 5, 2009} 
\begin{abstract} 
The tidal interaction of a (rotating or nonrotating) black hole with
nearby bodies produces changes in its mass, angular momentum, and
surface area.  Similarly, tidal forces acting on a Newtonian, viscous  
body do work on the body, change its angular momentum, and part of the
transferred gravitational energy is dissipated into heat. The
equations that describe the rate of change of the black-hole mass,
angular momentum, and surface area as a result of the tidal
interaction are compared with the equations that describe how the
tidal forces do work, torque, and produce heat in the Newtonian
body. The equations are strikingly similar, and unexpectedly, the
correspondence between the Newtonian-body and black-hole results is
revealed to hold in near-quantitative detail. The correspondence
involves the combination $k_2 \tau$ of ``Love quantities'' that
incorporate the details of the body's internal structure; $k_2$ is the
tidal Love number, and $\tau$ is the viscosity-produced delay between
the action of the tidal forces and the body's reaction.  The
combination $k_2 \tau$ is of order $GM/c^3$ for a black hole of mass
$M$; it does not vanish, in spite of the fact that $k_2$ is known to
vanish individually for a nonrotating black hole.    
\end{abstract} 
\pacs{04.20.-q, 04.25.Nx, 04.70.Bw}
\maketitle

\section{Introduction and summary} 

\subsection*{This work and its context} 

Flanagan and Hinderer have recently shown 
\cite{flanagan-hinderer:08, hinderer:08} that the tidal interaction of  
two neutron stars in the inspiral phase of their orbital evolution
leads to measurable effects in the gravitational waves emitted by the
binary system. Such a measurement can be used to constrain the radius
of each neutron star, and can thus reveal information about the
equation of state of nuclear matter. This exciting prospect has
provoked a flurry of activity that aims to advance our understanding
of tidal interactions between strongly self-gravitating bodies.  

Damour and Nagar \cite{damour-nagar:09}, and independently Binnington
and Poisson \cite{binnington-poisson:09}, have formulated a
relativistic theory of tidal Love numbers that applies to a neutron
star (or any other type of compact object) that is slightly deformed
(from a spherical shape) by tidal forces. A tidal Love number is a
dimensionless constant of proportionality that appears in the
relationship between the tidal forces and the induced body
deformation, measured by the mutipole moments of the mass 
distribution. Two types of tidal Love numbers occur in
general relativity: An electric-type Love number associated with
gravito-electric tidal forces, and a magnetic-type Love number
associated with gravito-magnetic tidal forces. In each case the Love 
number refers to the shared multipole order of the tidal field and the
induced deformation; there are two Love numbers for each
multipole order.  

The electric-type Love number is directly analogous to the tidal Love
number of Newtonian gravity (see, for example,
Ref.~\cite{murray-dermott:99}), and indeed, its relativistic
definition reduces to the Newtonian definition when the compactness $C
:= 2GM/(c^2 a)$ of the body is small. (Here $M$ denotes the body's
mass, and $a$ is its unperturbed radius; the compactness parameter 
$C \leq 1$ measures the importance of relativistic effects on the
body's internal structure.) The magnetic-type Love number, on the
other  hand, is a relativistic quantity that vanishes in the Newtonian
limit $C \to 0$.  The Binnington--Poisson theory
\cite{binnington-poisson:09} applies just as well to black holes, and
these authors have shown that the relativistic tidal Love numbers of a
nonrotating black hole are all zero.     

Other notions of Love numbers have been introduced in Newtonian
gravity \cite{murray-dermott:99}, and these also can be promoted to a
relativistic framework. For example, Damour and Nagar
\cite{damour-nagar:09} have introduced ``shape Love 
numbers'' in the relativistic theory of tidally deformed neutron
stars; these relate the tidal forces to the displacement of the stellar 
surface.  In a subsequent publication, Damour and Lecian
\cite{damour-lecian:09} extended the definition of shape Love numbers
to black holes, generalizing a previous analysis by Fang and Lovelace
\cite{fang-lovelace:05}. These works have contributed significantly to
our understanding of relativistic tidal interactions.   

My purpose in this paper is to contribute even further to this
understanding. I show that certain aspects of the tidal interaction of
black holes can be interpreted in purely Newtonian terms by exploiting
a beautiful analogy between the relativistic physics of black holes
and the Newtonian physics of viscous fluids. This analogy was first
noticed by Hartle \cite{hartle:73, hartle:74}, and it was more fully
fleshed out in the book {\it Black Holes: The Membrane Paradigm}
\cite{thorne-etal:86}. Here the analogy is pushed even further, and
unexpectedly, it is revealed to hold in near-quantitative detail. The
analogy involves an additional ``Love quantity'' beyond the tidal and
shape Love numbers described previously. The new quantity $\tau$ is
not dimensionless; it possesses the dimension of time, and in
Newtonian physics it corresponds to the viscosity-produced delay
between the action of the tidal forces and the body's reaction. I
shall refer to $\tau$ as the ``viscous delay.''   

\subsection*{Flux formulae for black holes} 

The physics to be interpreted concerns the rates at which a black hole
changes its mass, angular momentum, and surface area as a result of
the tidal interaction with external bodies. These were derived in 
Ref~\cite{poisson:04d, poisson:05}, and the flux formulae are
expressed in terms of quantities $\E_{ab}$ that provide a
characterization of the hole's tidal environment. This Cartesian
3-tensor is symmetric and tracefree (STF), and it represents the
quadrupole moment of the tidal field acting on the black hole. (Here
and below the multipole expansion of the tidal field is truncated to
the leading, quadrupole order; generalization to higher multipole
orders is possible.)  

In Newtonian theory (see, for example,
Ref.~\cite{mathis-poncin-lafitte:09} for a modern treatment), the
tidal-moment tensor is defined by first decomposing the Newtonian
potential as $U = U_{\rm body}  + U_{\rm ext}$, in which 
$U_{\rm body}$ is associated with the reference body and 
$U_{\rm ext}$ is produced by the external objects. We erect a
Cartesian coordinate system $x^a$ that is attached to the
center-of-mass of the reference body, and we define $\E_{ab}(t)  
:= -\partial_{ab} U_{\rm ext}$, in which the spatial derivatives of 
the external potential are evaluated at $x^a = 0$, the position of the
center-of-mass. Because $U_{\rm ext}$ satisfies Laplace's equation in
the body's neighborhood, $\E_{ab}$ is tracefree in addition to being
symmetric in its indices. 

We consider a context in which the tidal forces exerted by the
external bodies are weak, and in this case it is appropriate to
express the external potential as $U_{\rm ext} = U_0(t) + g_a(t) x^a  
- \frac{1}{2} \E_{ab}(t) x^a x^b + \cdots$, a Taylor expansion about
the body's center-of-mass. The spatially-constant term $U_0(t)$ is
irrelevant, and $g_a(t) := \partial_a U_{\rm ext}$ is responsible for
the body's acceleration in the field of the external objects; the
remaining terms are responsible for the tidal forces acting on the
body.   

In general relativity, the tidal moments $\E_{ab}(v) := C_{0a0b}$
are defined in terms of the components $C_{0a0b}$ of the spacetime
Weyl tensor evaluated in a region of spacetime that is far away from
the black hole, but still close relative to the external bodies. In
this case the indices refer to an inertial frame moving with the black
hole, and $v$ is an advanced-time coordinate on the event
horizon. (See Refs.~\cite{poisson:04d, poisson:05} for details.)      

For a nonrotating black hole of mass $M$, the rates at which the tidal
interaction changes the hole's mass and angular momentum are given by
Eqs.~(8.38) and (8.39) of Ref.~\cite{poisson:04d}; these are 
\begin{equation} 
\dot{M} = \frac{16}{45} M^6\dot{\E}^{ab} \dot{\E}_{ab} 
\label{eq:MdotS} 
\end{equation} 
and 
\begin{equation} 
\dot{J} = -\frac{32}{45} M^6 
\bigl( \epsilon^a_{\ cd} {\E}^{cb} s^d \bigr) \dot{\E}_{ab},
\label{eq:JdotS} 
\end{equation} 
respectively. Here an overdot indicates differentiation with respect
to advanced-time $v$, $\epsilon_{abc}$ is the Cartesian permutation
symbol, and $\dot{J} := \dot{J}_a s^a$ is the component of the vector
$\dot{J}^a$ in the direction of the (arbitrary) unit vector $s^a$. The
rate of change of the surface area is obtained from the first law of 
black-hole mechanics applied to a nonrotating black hole,
$(\kappa/8\pi) \dot{A} = \dot{M}$, where $\kappa = (4M)^{-1}$ is the 
hole's surface gravity. I employ relativistic units, so that $G = c =
1$.  

For a rapidly rotating black hole of mass $M$ and angular-momentum
$J$, the rate at which the angular momentum changes as a result of the
tidal interaction is given by Eq.~(9.39) of Ref.~\cite{poisson:04d};
this is 
\begin{eqnarray} 
\dot{J} &=& -\frac{16}{45} M^6 \Omega_H \Bigl( 1 + \sqrt{1-\chi^2}
\Bigr) \biggl[ 2(1 + 3\chi^2) \bigl( {\E}_{ab} {\cal E}^{ab} \bigr)
\nonumber \\ & & \mbox{} 
- 3\Bigl(1 + \frac{17}{4} \chi^2 \Bigr) 
   \bigl( {\E}_{ab} s^b {\E}^a_{\ c} s^c \bigr) 
\nonumber \\ & & \mbox{} 
+ \frac{15}{4} \chi^2 \bigl( {\cal E}_{ab} s^a s^b \bigr)^2 \biggr].
\label{eq:JdotK} 
\end{eqnarray} 
Here $\chi := J/M^2$ is the dimensionless Kerr parameter (which
is limited to the interval $0 \leq \chi < 1$), the unit vector $s^a$
is the direction of the hole's rotation axis, and 
\begin{equation} 
\Omega_H := \frac{\chi}{2M\bigl( 1 + \sqrt{1-\chi^2} \bigr)} 
\label{eq:Omega} 
\end{equation} 
is the angular velocity of the event horizon. An equation for
$\dot{M}$ is not available in this case, but at the level of accuracy
achieved by Eq.~(\ref{eq:JdotK}), $\dot{A}$ can still be obtained from
the first law: $(\kappa/8\pi) \dot{A} = -\Omega_H \dot{J}$, where
$\kappa = (2M)^{-1} \sqrt{1-\chi^2}(1-\sqrt{1-\chi^2})^{-1}$ is the
surface gravity of a Kerr black hole.  

Equations (\ref{eq:MdotS}), (\ref{eq:JdotS}), and (\ref{eq:JdotK}) 
differ from the original expressions displayed in
Ref.~\cite{poisson:04d} by terms involving the gravito-magnetic tidal
moments $\B_{ab}(v)$. These are switched off because they possess no
Newtonian analogues and cannot, therefore, be involved in a
correspondence with the tidal dynamics of a Newtonian body. The
neglect of the gravito-magnetic terms can be motivated on the basis of
a post-Newtonian approximation of the tidal fields
\cite{taylor-poisson:08}; in this context the terms in $\dot{M}$ and
$\dot{J}$ that involve $\B_{ab}$ are suppressed relative to those
involving $\E_{ab}$ by factors of order $(v_{\rm orb}/c)^2 \ll 1$, 
where $v_{\rm orb}$ measures the orbital velocity of the external
bodies.   

\subsection*{Tidal coupling of a Newtonian, viscous body} 

The key to a Newtonian interpretation of Eqs.~(\ref{eq:MdotS}),
(\ref{eq:JdotS}), and (\ref{eq:JdotK}) is the relation (Sec.~III D) 
\begin{equation} 
Q_{jk} = \frac{2}{3} n_2 a^5 \Omega_H^2 C_{jk} 
- \frac{2}{3} k_2 a^5 \Bigl( \E_{jk} - \tau \dot{\E}_{jk} \Bigr) 
\label{eq:Q_vs_E1}
\end{equation} 
that holds between the mass quadrupole moment $Q_{jk}$ of a Newtonian
body of mass $M$, unperturbed radius $a$, and angular velocity
$\Omega_H$, and the quadrupole moment $\E_{jk}$ of the applied tidal
field (an overdot indicates differentiation with respect to time
$t$). This relation is formulated in a frame $x^j$ that rotates
uniformly with angular velocity $\Omega_H$ relative to the global
inertial frame (the frame corotates with the body); this differs from
the original frame $x^a$, which moves with the body but does not
rotate. Equation (\ref{eq:Q_vs_E1}) involves the STF tensor 
$C_{jk} := \frac{1}{3} \delta_{jk} - s_j s_k$, where $s_j$ is a unit
vector that points in the direction of the rotation axis. It involves
also the dimensionless numbers $n_2$ (rotational Love number) and
$k_2$ (tidal Love number), and the viscous delay $\tau$. The body's
mass quadrupole moment is defined by 
$Q^{jk} = \int \rho (x^j x^k - \frac{1}{3} r^2 \delta^{jk})\, d^3x$,
in which $\rho$ is the mass density; this STF tensor vanishes when the
density profile is spherically symmetric.   

The first term on the right-hand side of Eq.~(\ref{eq:Q_vs_E1})
describes the body's rotational deformation (flattening of the poles,
bulging of the equator). The scalings with $\Omega_H^2$ and $a^5$ are
dictated by dimensional analysis, and the rotational Love number $n_2$
incorporates the details of the body's internal structure. The second
and third terms collectively describe the tidal deformation. Once more
the scaling with $a^5$ is dictated by dimensional analysis, and the
tidal Love number $k_2$ depends on the body's internal structure. The
third term is contributed by the body's internal viscosity, and
dimensional analysis reveals that $\tau \propto a\nu/M$, with $\nu$
denoting the (averaged) kinematic viscosity of the fluid. 

The relation of Eq.~(\ref{eq:Q_vs_E1}) relies on an assumption that
the time scale associated with changes in $\E_{jk}(t)$ is long
compared with the viscous delay $\tau$. In this context the second 
and third terms can be packaged approximately as 
$-\frac{2}{3} k_2 a^5 \E_{jk}(t-\tau)$, which reveals that viscosity
does indeed introduce a delay between the action of the
tidal forces (described by $\E_{jk}$) and the body's reaction
(measured by $Q_{jk}$). The viscous delay produces a misalignment
between the figure of the body deformation (described by $Q_{jk}$) and 
the figure of the tidal forces (measured by $\E_{jk}$). In the case in
which the tides are produced by a single external body, the tidal
bulge points in a direction that is not quite aligned with the
direction of the external body.      

Equation (\ref{eq:Q_vs_E1}) holds in the body's rotating frame $x^j$,
and $\dot{\E}_{jk}$ is the rate of change of the tidal moment as
perceived by an observer corotating with the body. In the nonrotating
frame $x^a$ the quadrupole moment is given instead by (Sec.~III E)  
\begin{equation} 
Q_{ab} = \frac{2}{3} n_2 a^5 \Omega_H^2 C_{ab} 
- \frac{2}{3} k_2 a^5 \Bigl( \E_{ab} - \tau \dot{\E}_{ab} 
- \tau \Delta \dot{\E}_{ab} \Bigr).  
\label{eq:Q_vs_E2}
\end{equation} 
In this expression $C_{ab} := \frac{1}{3} \delta_{ab} -
s_a s_b$, in which the unit vector $s_a$ continues to point in the
direction of the rotation axis; $\dot{\E}_{ab}$ is now the rate of  
change of the tidal moment as perceived by a nonrotating observer,
and the last term 
\begin{equation} 
\Delta \dot{\E}_{ab} := 2\Omega_H \epsilon_{cd(a} \E^c_{\ b)} s^d 
\label{eq:Delta_E} 
\end{equation}  
accounts for the rotation of the body frame relative to the global
inertial frame. 

\subsection*{Flux formulae for Newtonian body} 

The rate at which the tidal forces do work on the Newtonian body is
calculated as $\dot{W} = \frac{1}{2} Q_{ab} \dot{\E}^{ab}$ 
(Sec.~IV A), and substitution of Eq.~(\ref{eq:Q_vs_E2}) yields 
\begin{equation} 
\dot{W} = \frac{1}{3} (k_2 \tau) a^5\, \dot{\E}^{ab}  
\bigl( \dot{\E}_{ab} + \Delta \dot{\E}_{ab} \bigr). 
\label{eq:WdotN} 
\end{equation} 
Notice that $\dot{W}$ is proportional to $k_2 \tau$: No (net) work
is done unless the body is deformed by the tidal forces, and no (net)
work is done unless the deformation is delayed with respect to the
application of the forces. In the absence of viscosity, the tidal
bulge points in the direction of the external body, and the transfer
of gravitational energy from the tidal field to the body is fully
reversible and does not lead to a net performance of work. In the
presence of viscosity, on the other hand, the misalignment of the
tidal bulge allows for an irreversible transfer of energy from the
tidal field to the body, as described by Eq.~(\ref{eq:WdotN}).  

The torque exerted on the body by the tidal forces produces a change 
of angular momentum described by $\dot{J}_a = -\epsilon_{abc} 
Q^b_{\ p}\E^{p c}$ (Sec.~IV B). Substitution of Eq.~(\ref{eq:Q_vs_E2})
and projection along $s^a$ yields 
\begin{equation}  
\Omega_H \dot{J} = -\frac{1}{3} (k_2 \tau) a^5 \Delta\dot{\E}^{ab}  
\bigl( \dot{\E}_{ab}+ \Delta \dot{\E}_{ab} \bigr). 
\label{eq:JdotN} 
\end{equation}  
The result is multiplied by $\Omega_H$ to convert $\dot{J}$ into
$\dot{E}_{\rm rot} = \Omega_H \dot{J}$, the rate of change of the
body's rotational energy. Once more the result is proportional to $k_2
\tau$: The angular momentum changes if and only if the body is
deformed by the tidal forces and the deformation is delayed with
respect to the application of the forces. In the absence of viscosity,
the tidal bulge points in the direction of the external body, and the
tidal forces have no opportunity to exert a torque. In the presence of
viscosity, on the other hand, the misalignment of the tidal bulge
allows the tidal forces to exert a torque, and the end result is
described by Eq.~(\ref{eq:JdotN}).  

The rate at which viscosity generates heat can be determined from the 
energy-balance equation (Sec.~IV D) $\dot{Q} = \dot{W} - \Omega_H
\dot{J}$. Substitution of Eqs.~(\ref{eq:WdotN}) and (\ref{eq:JdotN})
yields 
\begin{equation} 
\dot{Q} = \frac{1}{3} (k_2 \tau) a^5 
\bigl( \dot{\E}^{ab} + \Delta \dot{\E}^{ab} \bigr)
\bigl( \dot{\E}_{ab} + \Delta \dot{\E}_{ab} \bigr). 
\label{eq:QdotN} 
\end{equation}  

\subsection*{Newtonian interpretation of black-hole fluxes}  

The black-hole and Newtonian-body results can be put in a close
correspondence if we associate the $\dot{W}$ of Eq.~(\ref{eq:WdotN})
with the $\dot{M}$ of Eq.~(\ref{eq:MdotS}), and the $\dot{J}$ of
Eq.~(\ref{eq:JdotN}) with the $\dot{J}$ of Eqs.~(\ref{eq:JdotS}) and
(\ref{eq:JdotK}). With these associations, the energy-balance equation
$\dot{Q} = \dot{W} - \Omega_H \dot{J}$ becomes the first law of
black-hole mechanics, $(\kappa/8\pi) \dot{A} = \dot{M} - \Omega_H
\dot{J}$, when we also associate $\dot{Q}$ with $(\kappa/8\pi)
\dot{A}$; this association is natural if we think of $\kappa/2\pi$ as
the black-hole temperature and $\frac{1}{4} A$ as its entropy.  

The nonrotating black hole corresponds to the case in which
$\dot{\E}_{ab}$ dominates over $\Delta \dot{\E}_{ab}$ in the Newtonian
equations; recall from Eq.~(\ref{eq:Delta_E}) that $\Delta
\dot{\E}_{ab}$ is proportional to $\Omega_H$ and therefore vanishes
when the body is nonrotating. In this case Eqs.~(\ref{eq:WdotN}) and
(\ref{eq:JdotN}) give 
\begin{equation} 
\dot{M} = \frac{1}{3} (k_2 \tau) a^5 \dot{\E}^{ab} \dot{\E}_{ab}, \quad 
\dot{J} = -\frac{2}{3} (k_2 \tau) a^5 \bigl( \epsilon^a_{\ cd} {\E}^{cb} s^d
\bigr) \dot{\E}_{ab}, 
\label{eq:comp_S} 
\end{equation} 
and these equations bear a striking resemblance 
with Eqs.~(\ref{eq:MdotS}) and (\ref{eq:JdotS}).  

The rapidly rotating black hole corresponds to the case in which
$\Delta \dot{\E}_{ab}$ dominates over $\dot{\E}_{ab}$; the changes in
the tidal moment are produced almost entirely by the body's own
rotation. In this case Eq.~(\ref{eq:JdotN}) gives 
$\Omega_H \dot{J} = -\frac{2}{3} (k_2 \tau) a^5 \Delta \dot{\E}^{ab} 
\Delta \dot{\E}_{ab}$, and expanding the tensorial expression produces  
\begin{equation} 
\dot{J} = -\frac{2}{3} (k_2 \tau) a^5 \Omega_H \Bigl[ 
2 \bigl( {\E}_{ab} {\cal E}^{ab} \bigr) 
- 3    \bigl( {\E}_{ab} s^b {\E}^a_{\ c} s^c \bigr) \Bigr].  
\end{equation} 
Comparison with Eq.~(\ref{eq:JdotK}) reveals another striking 
similarity, especially if we neglect the terms of order $\chi^2$ in
Eq.~(\ref{eq:JdotK}). This is appropriate, because according to
Eq.~(\ref{eq:Omega}), $\chi^2$ can be thought of as a relativistic
correction of order $(v_{\rm rot}/c)^2$, in which 
$v_{\rm rot} = a\Omega_H$ stands for the velocity of a fluid element
on the surface of the rotating body.  

Closer examination reveals that the correspondence produces agreement
between {\it all} numerical coefficients when we make the assignment   
\begin{equation} 
(k_2 \tau) a^5 = \frac{16}{15} \biggl( \frac{GM}{c^3} \biggr) 
\biggl( \frac{GM}{c^2} \biggr)^5 
\label{eq:corr} 
\end{equation}  
in the Newtonian equations. (I have restored the factors of $G$ and
$c$, which were previously set equal to unity). The 
equation indicates that as might be expected, the horizon length scale
$GM/c^2$ must play the role of the body radius $a$, while the horizon
time scale $GM/c^3$ must play the role of the effective viscous delay
$k_2 \tau$. This last result implies that the horizon can be assigned
an effective kinematic viscosity $k_2 \nu \sim GM/c$. 

The picture that emerges from this comparison is one in which the 
event horizon behaves as if it were a fictitious membrane of viscous
fluid; the fluid elements move with the speed of light, and the
streamlines are identified with the horizon's generators. This view,
of course, is compatible with the {\it membrane paradigm} of
black-hole physics \cite{thorne-etal:86}. 

The correspondence between the black-hole and Newtonian-body fluxes
depend on the combination $k_2 \tau$ of ``Love quantities,'' which is
constrained by Eq.~(\ref{eq:corr}); the constraint applies to
rapidly-rotating as well as nonrotating black holes, and it implies
that $k_2 \tau$ must be a finite (nonvanishing) quantity. It is known, 
however, that $k_2 = 0$ for a nonrotating black hole 
\cite{binnington-poisson:09}. One must therefore think of the
combination $k_2 \tau$ as being inseparable for (nonrotating) black
holes. While $k_2$ vanishes individually, $k_2 \tau$ is nevertheless a
finite quantity of order $GM/c^3$. It is not helpful to think of
$\tau$ as an individuated quantity that happens to be infinite.       

\subsection*{Organization of the paper} 

In the remaining sections of the paper I elaborate the Newtonian
theory of the interaction of a fluid body with a tidal environment
characterized by the tidal quadrupole moment $\E_{ab}(t)$. In addition
to establishing other interesting results, I provide derivations for
Eqs.~(\ref{eq:Q_vs_E1}), (\ref{eq:Q_vs_E2}), (\ref{eq:WdotN}),
(\ref{eq:JdotN}), and (\ref{eq:QdotN}).   

I begin in Sec.~II with a warmup exercise in which the fluid's
viscosity is set equal to zero; viscosity is introduced next in
Sec.~III. In Sec.~IV I calculate the rates at which the tidal forces
do work on the body, change its angular momentum, and lead to the
production of heat. For simplicity I perform all calculations assuming
that the body consists of an incompressible fluid; general results are
obtained from the special cases by inserting Love numbers where
appropriate.   

\section{Tidal interaction of a Newtonian body; nonviscous case}   

We consider a body of mass $M$, radius $a$, and angular velocity
$\Omega_H$ that would be spherical in isolation and in the absence of
rotation. The body is made of an incompressible fluid, it rotates
rigidly, and it is put in the presence of external objects that exert
tidal forces. As a result of its rotation and tidal interaction, the
body acquires a deformation. We calculate this deformation, first
assuming that the fluid's viscosity can be neglected. 

\subsection{Scales and assumptions} 

We already introduced the scaling quantities $M$, $a$, and
$\Omega_H$. The tidal interaction is characterized by an external mass
scale $M'$ and a distance scale $b$; it is measured by the tidal
moment tensor $\E_{jk}$, whose scale is given
by $\E \sim GM'/b^3$. The time scale associated with changes in the
tidal field, as viewed in the global inertial frame, is $\Omega^{-1}$,
where $\Omega$ is an angular velocity associated with the motion of
the external objects. In the body's rotating frame the time scale
becomes $|\Omega-\Omega_H|^{-1}$, and the tidal field is static if the
angular velocities match at all times. The tidal forces produce a
displacement of a fluid element at the body's surface; the length
scale of this displacement is $\xi \sim a^4\E/(GM)$, and we assume
that $\xi \ll a$. We also assume that the deformation produced by the
rotation is small, but we make no assumption regarding the relative
size of these effects.  

The velocity scale associated with fluid motions in the rotating frame
is $v \sim \xi |\Omega-\Omega_H|^{-1}$, or $v \sim a^4
|\Omega-\Omega_H|^{-1}\E /(GM)$. We assume that $v$ is much smaller
than the linear velocity $a\Omega_H$ of a fluid element following the
body's rotation  (as viewed in the global inertial frame). The
inequality $v \ll a\Omega_H$ implies that the Coriolis effect can be
ignored, but centrifugal terms will make an appearance in the fluid
equations. This assumption constrains the size of the tidal
interaction. Assuming that $\Omega_H$ is larger than or at least
comparable to $\Omega$, the inequality translates to 
$\E \ll GM/a^3$, or $(a/b)^3 \ll M/M'$; the body must be well
separated from the objects that produce the tidal forces. As an
additional assumption we take $v |\Omega-\Omega_H|^{-1}$, the velocity
scale divided by the  tidal time scale,  to be much smaller than 
$\E a$, the scale associated with the tidal forces (per unit mass)
acting within the body. This inequality allows us to neglect the
inertial term $\partial_t v$ in the fluid equation. It translates to
$(\Omega-\Omega_H)^2 \ll GM/a^3$, and the assumption constrains the  
size of the angular velocities; they must be small compared with the
Keplerian angular velocity of a particle at the body's surface. All
the assumptions made here are reasonable, and they do not severely
restrict the range of physical situations that can be considered. 

\subsection{Fluid equations; nonrotating frame} 

We first formulate the fluid equations in a frame $x^{\bar{a}}$ that
moves with the body's center-of-mass but does not rotate relative to 
the global inertial frame. (This frame was denoted $x^a$ in Sec.~I,
and here we put an overbar on the index to increase notational
clarity.) The frame $x^{\bar{a}}$ is translated by $r^{\bar{a}}(t)$ 
relative to the origin of the global inertial frame; this is the
position of the body's center-of-mass.  

The equations of fluid dynamics are reviewed, for example, in Secs.~1 
and 2 of Ref.~\cite{landau-lifshitz:b6}. A perfect fluid is governed
by (i) the continuity equation  
\begin{equation} 
\partial_t \rho +\partial_{\bar{a}} (\rho v^{\bar{a}}) = 0,  
\end{equation} 
in which $\rho$ is the mass density and $v^{\bar{a}}$ the velocity
field; (ii) Euler's equation 
\begin{equation} 
\rho \frac{d v^{\bar{a}}}{dt} = -\partial_{\bar{a}} p 
+ \rho \partial_{\bar{a}} U - \rho \ddot{r}_{\bar{a}},  
\end{equation} 
in which $d v^{\bar{a}}/dt = \partial_t v^{\bar{a}} 
+ v^{\bar{b}} \partial_{\bar{b}} v^{\bar{a}}$ is the advective time
derivative, $p$ the pressure, $U$ the gravitational potential, and the
last term is an inertial force density associated with the translation
of the frame $x^{\bar{a}}$ relative to the global inertial frame;
(iii) Poisson's equation 
\begin{equation} 
\nabla^2 U = -4\pi G \rho 
\end{equation} 
for the gravitational potential; and (iv) an equation of state that
relates the pressure to the density. 

In the case of an incompressible fluid, $\rho = \mbox{constant}$ and
the fluid equations become 
\begin{equation} 
\partial_{\bar{a}} v^{\bar{a}} = 0
\end{equation} 
and 
\begin{equation} 
\frac{dv^{\bar{a}}}{dt} = \partial_{\bar{a}} \Bigl( -p/\rho  
+ U - \ddot{r}_{\bar{b}} x^{\bar{b}} \Bigr).  
\end{equation} 
In this case the pressure is unrelated to the density. 

It is useful to decompose the Newtonian potential as 
$U = U_{\rm body} + U_{\rm ext}$, in which $U_{\rm body}$ is
associated with the body and $U_{\rm ext}$ is produced by the external
objects. We also decompose the external potential as 
$U_{\rm ext}(t,x^{\bar{a}}) = U_0(t) + g_{\bar{a}}(t) x^{\bar{a}} 
+ U_{\rm tidal}(t,x^{\bar{a}})$, in which $U_0(t) 
:= U_{\rm ext}(t,r^{\bar{a}})$ and 
$g_{\bar{a}} := \partial_{\bar{a}} U_{\rm ext}(t,r^{\bar{a}})$; this  
provides a definition for the tidal potential $U_{\rm tidal}$. Because
the body moves according to $\ddot{r}_{\bar{a}} = g_{\bar{a}}$, the
final form of Euler's equation is 
\begin{equation} 
\frac{dv^{\bar{a}}}{dt} = \partial_{\bar{a}} \bigl( -p/\rho  
+ U_{\rm body} + U_{\rm tidal} \bigr).  
\end{equation} 

\subsection{Fluid equations; rotating frame} 

The transformation from the frame $x^{\bar{a}} =
[\bar{x},\bar{y},\bar{z}]$ to the rotating frame $x^j = [x,y,z]$ is
described by  
\begin{subequations}
\begin{align}  
x &= \bar{x} \cos\Omega_H t + \bar{y} \sin\Omega_H t, \\
y &= -\bar{x} \sin\Omega_H t + \bar{y} \cos\Omega_H t, \\ 
z &= \bar{z}, 
\end{align}
\end{subequations} 
if we take the rotation axis to coincide with the $z$-axis. The
transformation can be written more formally as 
\begin{equation} 
x^j = \Lambda^j_{\ \bar{a}} x^{\bar{a}}.  
\end{equation} 
The velocity vector transforms as 
\begin{subequations}
\begin{align}  
v^x &= v^{\bar{x}} \cos\Omega_H t + v^{\bar{y}} \sin\Omega_H t
\nonumber \\ & \qquad \mbox{} 
 + \Omega_H \bigl(-\bar{x} \sin\Omega_H t 
  + \bar{y} \cos\Omega_H t \bigr), \\ 
v^y &=-v^{\bar{x}} \sin\Omega_H t + v^{\bar{y}} \cos\Omega_H t
\nonumber \\ & \qquad \mbox{} 
  - \Omega_H \bigl(\bar{x} \cos\Omega_H t 
  + \bar{y} \sin\Omega_H t \bigr),  \\
v^z &= v^{\bar{z}}, 
\end{align}
\end{subequations} 
which can be written formally as 
\begin{equation} 
v^j = \Lambda^j_{\ \bar{a}} (v^{\bar{a}} +\Delta v^{\bar{a}}), 
\end{equation} 
where 
\begin{equation} 
\Delta v^{\bar{a}} := \Gamma^{\bar{a}}_{\ \bar{b}} x^{\bar{b}} 
\end{equation} 
accounts for the time-dependence of the transformation. The matrix
$\Gamma^{\bar{a}}_{\ \bar{b}}$ is formally
defined by $\Gamma^{\bar{a}}_{\ \bar{b}} = \Lambda^{\bar{a}}_{\ j}
\dot{\Lambda}^j_{\ \bar{b}}$, where $\Lambda^{\bar{a}}_{\ j}$ is the
inverse of the matrix $\Lambda^j_{\ \bar{a}}$. Calculation reveals that 
\begin{equation} 
 \Gamma^{\bar{a}}_{\ \bar{b}} = \Omega_H 
\epsilon^{\bar{a}}_{\ \bar{b}\bar{c}} s^{\bar{c}}, 
\end{equation} 
where $\epsilon_{\bar{a}\bar{b}\bar{c}}$ is the permutation symbol and 
the vector $s^{\bar{c}} = [0,0,1]$ denotes the direction of the
rotation axis.  

Making the substitution in Euler's equation produces additional
terms that originate from the noninertial nature of the
transformation. The first collection of additional terms are linear in
both $\Omega_H$ and $v^j$, and are responsible for the Coriolis
effect; these we neglect, according to the assumptions formulated in
Sec.~II A. The second collection of terms are quadratic in $\Omega_H$ 
and independent of $v^j$, and give rise to centrifugal repulsion;
these we keep. To simplify the equation further we neglect all terms
quadratic in $v^j$ and the inertial term $\partial_t v^j$, again in
accordance with the stated assumptions. 

As a result of these substitutions we find that the fluid equations in
the rotating frame are $\partial_j v^j = 0$,
\begin{equation} 
\partial_j \bigl( -p/\rho + U_{\rm body} + U_{\rm tidal} 
+ \Omega_H^2 C \bigr) = 0, 
\label{eq:Euler_rot} 
\end{equation} 
and the gravitational potentials are determined by 
$\nabla^2 U_{\rm body} = -4\pi G\rho$ and $\nabla^2 U_{\rm tidal} 
= 0$. Here $\Omega_H^2 C = \frac{1}{2}\Omega_H^2 (x^2 + y^2)$ is the
centrifugal potential. It is useful to express $C$ as 
\begin{equation} 
C = \frac{1}{6} r^2 + \frac{1}{2} C_{jk} x^j x^k, 
\label{eq:centrif} 
\end{equation} 
where 
\begin{equation} 
C_{jk} = \frac{1}{3} \delta_{jk} - s_j s_k 
\end{equation} 
is a symmetric-tracefree (STF) tensor. The first term in $C$'s
decomposition is spherically symmetric; the second has a
quadrupolar ($\ell = 2$) structure. 

\subsection{Tidal field} 

The gravitational potential was decomposed as 
\begin{equation} 
U = U_{\rm body} + U_{\rm tidal}, 
\label{eq:Utotal} 
\end{equation} 
with $U_{\rm body}$ representing the potential of the (deformed) body, 
and $U_{\rm tidal}$ representing the tidal potential. To leading order
in a Taylor expansion of the external potential about the body's
center-of-mass, this is  
\begin{equation} 
U_{\rm tidal} = -\frac{1}{2} \E_{jk}(t) x^j x^k, 
\label{eq:Utidal} 
\end{equation} 
where the tidal moment $\E_{jk}$ is a time-dependent STF tensor
that does not depend on spatial position. The tidal potential also has
a quadrupolar structure. 

A specific example of a tidal field is one produced by an external
object of mass $M'$ on a circular orbit of radius $b$ around the
reference body. If we let $m^j$ be a unit vector that points in the
direction of the external object (from the body's center-of-mass), a
simple computation reveals that the tidal moment is given by  
\begin{equation} 
\E_{jk}(t) = -\frac{GM'}{3b^3} \Bigl( m_j m_k 
- \frac{1}{3} \delta_{jk} \Bigr). 
\label{eq:E_example}
\end{equation} 
If we take the external object to move in the equatorial plane of the
rotating body, the unit vector is given by 
\begin{equation} 
m^j(t) = [\cos(\Omega-\Omega_H)t, \sin(\Omega-\Omega_H)t,0]
\end{equation} 
in the rotating frame, where $\Omega$ is the orbital angular velocity. 

\subsection{Body deformation} 

The tidal forces and body rotation produce a deformation in the body's
shape, which can be described by 
\begin{equation} 
R = a \bigl( 1 + e_{jk} n^j n^k \bigr), 
\end{equation} 
where $n^j = x^j/r$ is the radial unit vector. The equation $r =
R(n^j)$ describes the position of the body's deformed surface. We let
the deformation assume a quadrupolar shape, in accordance with the
fact that each deforming force is quadrupolar in nature; $e_{jk}$ is 
a STF tensor that must be determined by solving the fluid equations. 

\subsection{Body potential} 

The body potential is given by 
\begin{equation} 
U_{\rm body, out} = \frac{GM}{r} 
+ \frac{3}{2} G Q_{jk} \frac{x^j x^k}{r^5} 
\end{equation} 
outside the body, and 
\begin{equation} 
U_{\rm body, in} = U_0(r) + U_{jk} x^j x^k 
\label{eq:Uin} 
\end{equation} 
inside the body. In each case the deviation from a spherical potential
is measured by a STF tensor; we have the quadrupole-moment tensor
$Q_{jk}$ outside, and $U_{jk}$ inside. 

The body mass is related to the density by
\begin{equation} 
M = \frac{4\pi}{3} \rho a^3. 
\end{equation} 
The quadrupole-moment tensor is determined by evaluating $Q^{jk} =
\int \rho x^{\stf{jk}}\, d^3x$, where $x^{\stf{jk}} := x^j x^k -
\frac{1}{3} r^2 \delta^{jk}$; a quick computation returns 
\begin{equation} 
Q_{jk} = \frac{2}{5} M a^2 e_{jk}. 
\end{equation} 
The internal potential $U_0$ is obtained by solving Poisson's equation
for a spherical configuration of constant density $\rho$ and radius
$a$; the result is 
\begin{equation} 
U_0(r) = \frac{GM}{2a} \bigl( 3 - r^2/a^2 ).
\end{equation} 
And finally, $U_{jk}$ is obtained by demanding continuity of the body
potential across the deformed surface; this yields 
\begin{equation} 
U_{jk} = \frac{3}{5} \frac{GM}{a^3} e_{jk}. 
\end{equation} 

\subsection{Pressure} 

The pressure field inside the body is expressed as 
\begin{equation} 
p/\rho = p_0(r)/\rho + p_{jk} x^j x^k, 
\label{eq:pressure} 
\end{equation}
where $p_{jk}$ is another STF tensor. This form is motivated by the
fact that according to Eq.~(\ref{eq:Euler_rot}), the pressure must
satisfy the Poisson equation $\nabla^2 (p/\rho) = -4\pi G\rho 
+ \Omega_H^2$. Because the right-hand side is spherically symmetric,
$p_0/\rho$ must satisfy this equation by itself, and the additional
term must be a solution to Laplace's equation; a quadrupolar structure 
necessarily selects the assumed form $p_{jk} x^j x^k$. 

The spherically-symmetric component of the pressure is easily obtained
by integrating Poisson's equation; we get 
\begin{equation} 
p_0(r)/\rho = \frac{GM}{2a} \biggl( 1 
- \frac{\Omega_H^2}{4\pi G \rho} \biggr) \Bigl( 1 - r^2/a^2 \Bigr) 
\end{equation}  
when we demand that the unperturbed pressure vanish on the unperturbed
surface $r=a$. An expression for $p_{jk}$ is obtained when we demand
that the perturbed pressure vanish on the perturbed surface. If we
let $\xi := R - a = a e_{jk} n^j n^k$ describe the deformation, we
have that $0 = p(a+\xi) = p_0(a) + ap'_0(a) e_{jk} n^j n^k 
+ a^2 p_{jk} n^j n^k$, in which a prime indicates differentiation with
respect to $r$. The computation returns 
\begin{equation} 
p_{jk} =\frac{GM}{a^3} e_{jk} 
\end{equation} 
after neglecting the correction of order 
$\Omega_H^2/(4\pi G\rho)$ in $p_0(r)$.  

\subsection{Solution} 

Euler's equation (\ref{eq:Euler_rot}) produces the constraint 
\begin{equation} 
-p_{jk} + U_{jk} - \frac{1}{2} \E_{jk} + \frac{1}{2} \Omega_H^2 C_{jk}
= 0 
\end{equation} 
when we insert Eqs.~(\ref{eq:centrif}), (\ref{eq:Utidal}),
(\ref{eq:Uin}), and (\ref{eq:pressure}) into it. Since $p_{jk}$ and
$U_{jk}$ are both proportional to the deformation tensor $e_{jk}$, we
quickly deduce that 
\begin{equation}
e_{jk} = \frac{5}{4} \frac{a^3}{GM}\bigl( \Omega^2_H C_{jk} 
- \E_{jk} \bigr). 
\end{equation} 
This result informs us that the deformation is a combination of
rotational and tidal effects, as expected. We also have 
\begin{eqnarray} 
p_{jk} &=& \frac{5}{4} \bigl( \Omega^2_H C_{jk} - \E_{jk} \bigr), \\ 
U_{jk} &=& \frac{3}{4}\bigl( \Omega^2_H C_{jk} - \E_{jk} \bigr),  
\end{eqnarray} 
and the quadrupole-moment tensor is 
\begin{equation} 
Q_{jk} = \frac{1}{2} \frac{a^5}{G} \bigl( \Omega^2_H C_{jk} 
- \E_{jk} \bigr).
\end{equation} 

\section{Tidal interaction of a Newtonian body; viscous case}  

We now incorporate viscosity into the fluid model. The fluid's shear
viscosity is measured by $\nu$, the coefficient of kinematic
viscosity, which we assume to be uniform within the fluid.   

\subsection{Scales and assumptions} 

We continue to deal with the scaling quantities introduced in Sec.~II A,
and with one exception, we make the same assumptions regarding them as 
we did in the preceding section. The only change concerns the time
scale of the problem, which previously was provided by the orbital
dynamics as viewed in the rotating frame; the only relevant time scale
was $|\Omega-\Omega_H|^{-1}$. The addition of viscosity introduces
another relevant time scale, namely the delay between the action of
the applied tidal force and the body's response. As we shall see, the
time scale for the viscous delay is $\tau \sim a\nu/(GM)$, and we
assume that this is much shorter than the orbital time scale: 
$\tau \ll |\Omega-\Omega_H|^{-1}$; this inequality constrains the size
of the kinematic viscosity.   

The change of time scale has repercussions on the velocity scale. It
is now given by $v \sim \xi/\tau$, where $\xi \sim a^4 \E/(GM)$ is
still the length scale associated with fluid displacements. To
neglect the Coriolis effect with still demand that $v \ll a\Omega_H$,
but the inequality now translates to $\E \ll (GM/a^3)(\Omega_H \tau)$,
or $(a/b)^3 \ll (M/M') (\Omega_H \tau)$; since $\Omega_H\tau \ll 1$ we
find that the bodies must be even more widely separated. To neglect
the inertial term in the fluid equation we now demand that $v/\tau \ll
\E a$, and this translates to $\tau \gg \sqrt{a^3/(GM)}$, another
constraint on the kinematic viscosity. (The constraints are compatible
with each other whenever $|\Omega - \Omega_H| \ll \tau^{-1} \ll
\Omega_K$, where $\Omega_K = \sqrt{GM/a^3}$ is the Keplerian angular
velocity of a particle at the body's surface.) 

\subsection{Fluid equations} 

Under the stated assumptions, the fluid equations in the rotating
frame are now (see, for example, Sec.~15 of
Ref.~\cite{landau-lifshitz:b6})  
\begin{equation} 
\partial_j v^j = 0
\end{equation} 
and 
\begin{equation} 
\partial_j \bigl( -p/\rho + U_{\rm body} + U_{\rm tidal} 
+ C \bigr) + \nu \nabla^2 v^j = 0,  
\label{eq:NavierStokes} 
\end{equation} 
with the last term giving rise to all effects associated with the
viscosity. The centrifugal potential is still given by  
\begin{equation} 
C = \frac{1}{6} \Omega_H^2 r^2 
+ \frac{1}{2} \Omega_H^2 C_{jk} x^j x^k, 
\end{equation} 
and the internal gravitational potential is still given by 
\begin{equation} 
U_{\rm body} + U_{\rm tidal} = U_0(r) + U_{jk} x^j x^k 
- \frac{1}{2} \E_{jk} x^j x^k, 
\end{equation} 
with $U_0(r) = GM( 3 - r^2/a^2 )/(2a)$ and $U_{jk} = 3GM 
e_{jk}/(5a^3)$. The body deformation is still described by 
$R = a(1 + e_{jk} n^j n^k)$, and the quadrupole moment is still
$Q_{jk} = \frac{2}{5} M a^2 e_{jk}$.  
  
The presence of viscosity affects the nature of the boundary
conditions at the body's surface. Instead of the no-pressure condition
imposed previously, we must now impose the no-normal-stress condition 
\begin{equation} 
p n_j - \rho\nu \bigl(\partial_j v_k + \partial_k v_j \bigr) n^k = 0
\end{equation} 
at $r = R$. In addition, we must relate the normal component of the
velocity vector to the rate of change of the surface displacement: 
\begin{equation} 
v_j n^j \Bigr|_{r=a} = \partial_t R. 
\end{equation} 
Unlike the preceding equation (which involves the unperturbed piece of
the pressure), this equation can be formulated at $r=a$ instead of
$r=R$, because it involves perturbed quantities only. 

Because the pressure still satisfies the Poisson equation 
$\nabla^2 (p/\rho) = -4\pi G\rho + \Omega_H^2$, we may still express
it as 
\begin{equation} 
p/\rho = p_0(r)/\rho + p_{jk} x^j x^k. 
\end{equation} 
The expression for $p_0(r)$ is unchanged, but $p_{jk}$ is no longer
equal to $GM e_{jk} /a^3$; its new value will be determined by the new
boundary conditions. Using the available information, we find that
these become 
\begin{equation} 
\nu \bigl(\partial_j v_k + \partial_k v_j \bigr) n^k 
= a^2 n_j \Bigl( p_{kn} - \frac{GM}{a^3} e_{kn} \Bigr) n^k n^n 
\label{eq:bound1} 
\end{equation} 
and 
\begin{equation} 
v_j n^j = a \dot{e}_{jk} n^j n^k. 
\label{eq:bound2} 
\end{equation} 
Both equations are now formulated at $r=a$. 

\subsection{Velocity field} 

At this stage the unknowns are the velocity field $v^j$, the
displacement tensor $e_{jk}$, and the pressure tensor $p_{jk}$; these 
must all be determined in terms of $C_{jk}$ and $\E_{jk}$. The
velocity field obeys a Poisson equation with a quadrupolar source
term. The solution must be a $\ell = 2$ vectorial harmonic, and it can
therefore be expressed as a linear superposition of terms 
$r^2 V_{jk} x^k$, $V_{kn} x^j x^k x^n$, and $V_{jk} x^k$, with
$V_{jk}$ a time-dependent STF tensor; the superposition is constrained
by the divergence-free condition $\partial_j v^j = 0$. The boundary
condition of Eq.~(\ref{eq:bound2}) relates $V_{jk}$ to $\dot{e}_{jk}$,
and Eq.~(\ref{eq:bound1}) determines $p_{jk}$ in terms of $e_{jk}$ and
its time derivative. Finally, the Navier-Stokes equation
(\ref{eq:NavierStokes}) gives rise to a differential equation for the
displacement tensor $e_{jk}$.   

After going through these manipulations and sorting out the algebra,
we arrive at the differential equation 
\begin{equation} 
\tau \dot{e}_{jk} + e_{jk} = \frac{5}{4} \frac{a^3}{GM} 
\bigl( \Omega^2_H C_{jk} - \E_{jk} \bigr), 
\label{eq:edot} 
\end{equation} 
where 
\begin{equation} 
\tau := \frac{19}{2} \frac{a\nu}{GM} 
\end{equation} 
is the viscous delay. We also obtain the pressure tensor  
\begin{equation} 
p_{jk} = \frac{GM}{a^3} \Bigl( e_{jk} 
- \frac{4\tau}{95} \dot{e}_{jk} \Bigr) 
\end{equation} 
and the velocity field 
\begin{equation} 
v^j = -\frac{1}{5a^2} \Bigl[ (5r^2-8a^2) \dot{e}_{jk} x^k 
- 2 \dot{e}_{kn} x^j x^k x^n \Bigr]. 
\end{equation} 

\subsection{Solution} 

The solution to the differential equation $\tau \dot{e} + e = f$ is 
\begin{equation} 
e(t) = e^{-t/\tau} \Bigl[ e(0) 
+ \frac{1}{\tau} \int_0^t e^{t'/\tau} f(t')\, dt'\Bigr]. 
\end{equation} 
Two integrations by parts bring this to the form 
\begin{eqnarray} 
e(t) &=& \bigl[ e(0) - f(0) + \tau \dot{f}(0) \bigr] e^{-t/\tau} 
+ f(t) - \tau \dot{f}(t)
\nonumber \\ & & \mbox{} 
+ \tau e^{-t/\tau} \int_0^t e^{t'/\tau} \ddot{f}(t')\, dt'.
\end{eqnarray} 
At times $t \gg \tau$ we may neglect the first group of terms, and we 
may also neglect the integral if $f(t)$ varies on a time scale that is long
compared with $\tau$. Under these circumstances the solution is 
\begin{equation} 
e(t) = f(t) -\tau \dot{f}(t) + O\bigl[ e(0) e^{-t/\tau} \bigr] 
+ O\bigl[\tau^2 \ddot{f}(t) \bigr]. 
\end{equation} 
Within the stated error, this is $e(t) = f(t-\tau)$. We see that the
displacement at time $t$ is related to the behavior of the driving
force at the earlier time $t-\tau$, and the delay is given precisely
by $\tau$. The viscosity therefore introduces a phase lag between the
applied force and the response, and as we shall see, this dephasing
turns out to be crucial for the physics of tidal work, torque, and
heating.   

In our context the role of the driving force $f$ is played by the
tidal moment $\E_{jk}$, which varies on a time scale of order
$|\Omega - \Omega_H|^{-1}$. We already have stated our assumption that 
this time scale is much longer than $\tau$. In these circumstances,
and ignoring the transients that decay exponentially, the solution to 
Eq.~(\ref{eq:edot}) is 
\begin{equation} 
e_{jk} = \frac{5}{4} \frac{a^3}{GM}\bigl( \Omega_H^2 C_{jk} 
- \E_{jk} + \tau \dot{\E}_{jk} \bigr). 
\end{equation} 
With this we find that 
\begin{eqnarray} 
p_{jk} &=& \frac{5}{4} \Bigl( \Omega_H^2 C_{jk} 
- \E_{jk} + \frac{99\tau}{95} \dot{\E}_{jk} \Bigr), \\
U_{jk} &=& \frac{3}{4}\Bigl( \Omega_H^2 C_{jk} 
- \E_{jk} + \tau \dot{\E}_{jk} \Bigr), 
\end{eqnarray} 
and 
\begin{equation} 
v^j = \frac{a}{4GM} \Bigl[ (5r^2-8a^2) \dot{\E}_{jk} x^k 
- 2 \dot{\E}_{kn} x^j x^k x^n \Bigr]. 
\label{eq:v_field} 
\end{equation} 
The quadrupole moment tensor is now 
\begin{equation} 
Q_{jk} = \frac{a^5}{2G} \Bigl( \Omega_H^2 C_{jk} 
- \E_{jk} + \tau \dot{\E}_{jk} \Bigr). 
\label{eq:Qrot} 
\end{equation} 

\subsection{Transformation to the nonrotating frame} 

For later purposes we transform our results from the rotating frame
$x^j$ back to the nonrotating frame $x^{\bar{a}}$; the transformation
is described by $x^{\bar{a}} = \Lambda^{\bar{a}}_{\ j} x^j$, and it is
the inverse of the one considered in Sec.~II C. 

As we have seen, the velocity field transforms as 
\begin{equation} 
v^{\bar{a}} = \Lambda^{\bar{a}}_{\ j} ( v^j + \Delta v^j),
\end{equation}
where 
\begin{equation} 
\Delta v^j := \Gamma^j_{\ k} x^k
\end{equation} 
accounts for the time-dependence of the transformation; here 
\begin{equation} 
\Gamma^j_{\ k} := \Lambda^j_{\ \bar{a}} \dot{\Lambda}^{\bar{a}}_{\ k}  
= -\Omega_H \epsilon^j_{\ kn} s^n, 
\end{equation} 
with $s^n = [0,0,1]$ denoting the direction of the rotation axis. 

In the nonrotating frame the tidal moment is given by 
\begin{equation} 
\E_{\bar{a}\bar{b}} = \Lambda^j_{\ \bar{a}} \Lambda^k_{\ \bar{b}}
\E_{jk}. 
\end{equation} 
The transformation of its time derivative, however, must account for   
the time-dependence of the transformation. Here we have 
\begin{equation} 
\dot{\E}_{\bar{a}\bar{b}} = \Lambda^j_{\ \bar{a}} \Lambda^k_{\ \bar{b}} 
\bigl( \dot{\E}_{jk}- \Delta \dot{\E}_{jk} \bigr)
= \Lambda^j_{\ \bar{a}} \Lambda^k_{\ \bar{b}}\, \dot{\E}_{jk} 
- \Delta \dot{\E}_{\bar{a}\bar{b}}, 
\label{eq:Edot_transf} 
\end{equation} 
with
\begin{equation} 
\Delta \dot{\E}_{jk} := \Gamma^n_{\ j} \E_{nk} 
+ \Gamma^n_{\ k} \E_{jn} 
= 2\Omega_H \epsilon_{pq(j} \E^p_{\ k)} s^q 
\end{equation} 
and 
\begin{equation} 
\Delta \dot{\E}_{\bar{a}\bar{b}} := 
\Lambda^j_{\ \bar{a}} \Lambda^k_{\ \bar{b}} \Delta \dot{\E}_{jk}
= 2\Omega_H \epsilon_{\bar{c}\bar{d}(\bar{a}} 
  \E^{\bar{c}}_{\ \bar{b})} s^{\bar{d}}.   
\end{equation} 

The transformation of the quadrupole-moment tensor is
$Q_{\bar{a}\bar{b}} = \Lambda^j_{\ \bar{a}} \Lambda^k_{\ \bar{b}}
Q_{jk}$. With Eqs.~(\ref{eq:Qrot}) and (\ref{eq:Edot_transf}), this is 
\begin{equation} 
Q_{\bar{a}\bar{b}} = \frac{a^5}{2G} \Bigl( 
\Omega_H^2 C_{\bar{a}\bar{b}}  
- \E_{\bar{a}\bar{b}} + \tau \dot{\E}_{\bar{a}\bar{b}} 
+ \tau \Delta \dot{\E}_{\bar{a}\bar{b}} \Bigr), 
\label{eq:Q_inertial} 
\end{equation} 
with the last term accounting for the effect of the body's rotation on 
the tidally-induced quadrupole moment.  

\subsection{Example} 

To illustrate the way in which the preceding (simple, but potentially
confusing) tensorial transformations work, we examine the case in 
which the rotating-frame tidal moment is given by 
Eq.~(\ref{eq:E_example}), 
\begin{equation} 
\E_{jk} = -\frac{GM'}{3b^3} \Bigl( m_j m_k 
- \frac{1}{3} \delta_{jk} \Bigr), 
\end{equation}
with $m^j = [\cos(\Omega-\Omega_H)t, \sin(\Omega-\Omega_H)t, 0]$. 
The transformed tidal moment is  
\begin{equation}
\E_{\bar{a}\bar{b}} = -\frac{GM'}{3b^3} \Bigl( m_{\bar{a}} m_{\bar{b}}  
- \frac{1}{3} \delta_{\bar{a}\bar{b}} \Bigr),
\end{equation}
with $m^{\bar{a}} = [\cos\Omega t, \sin\Omega t,0]$. The
time-derivative of the rotating-frame tidal field is 
\begin{equation}
\dot{\E}_{jk} =-\frac{GM'}{3b^3} (\Omega-\Omega_H) 
\bigl( m_j \phi_k + \phi_j m_k \bigr),
\end{equation}
where $\phi^j := [-\sin(\Omega-\Omega_H)t, \cos(\Omega-\Omega_H)t, 
0]$. From this we find that 
\begin{equation}
\Lambda^j_{\ \bar{a}} \Lambda^k_{\ \bar{b}} \dot{\E}_{jk}  
= -\frac{GM'}{3 b^3} (\Omega-\Omega_H) 
\bigl( m_{\bar{a}} \phi_{\bar{b}} + \phi_{\bar{a}} m_{\bar{b}} \bigr),
\end{equation}
with $\phi^{\bar{a}} = [-\sin\Omega t, \cos\Omega t,0]$. On the other 
hand,  
\begin{equation}
\Delta \dot{\E}_{jk} = \frac{GM'}{3 b^3} \Omega_H 
\bigl( m_j \phi_k + \phi_j m_k \bigr) 
\end{equation}
and 
\begin{equation}
\Delta \dot{\E}_{\bar{a}\bar{b}} =\frac{GM'}{3 b^3} \Omega_H 
\bigl( m_{\bar{a}} \phi_{\bar{b}} + \phi_{\bar{a}} m_{\bar{b}} \bigr).
\end{equation}
We see that these pieces add up correctly to produce the expected 
\begin{equation} 
\dot{\E}_{\bar{a}\bar{b}} = -\frac{GM'}{3 b^3} \Omega 
\bigl( m_{\bar{a}} \phi_{\bar{b}} + \phi_{\bar{a}} m_{\bar{b}} \bigr).
\end{equation}
The combination of terms that appears in $Q_{\bar{a}\bar{b}}$,
however, is 
\begin{equation}
\dot{\E}_{\bar{a}\bar{b}} + \Delta \dot{\E}_{\bar{a}\bar{b}} 
= -\frac{GM'}{3 b^3} (\Omega - \Omega_H) 
\bigl( m_{\bar{a}} \phi_{\bar{b}} + \phi_{\bar{a}} m_{\bar{b}} \bigr), 
\end{equation}
and we see that this vanishes when the external object is corotating
with the body. 

\section{Tidal work, torque, and heating} 

\subsection{Tidal work} 

The rate at which the tidal forces do work on the rotating body is
calculated as 
\[
\dot{W} = \int \bigl(-\rho \E_{\bar{a}\bar{b}} x^{\bar b} \bigr) 
v^{\bar{a}}\, d^3\bar{x}. 
\]
This is the integral of the tidal force density $-\rho
\E_{\bar{a}\bar{b}} x^{\bar b}$ times $v^{\bar{a}}$, the rate of
fluid displacement. This is equal to 
\begin{equation} 
\dot{W} = -\frac{1}{2} \E_{\bar{a}\bar{b}} \dot{Q}^{\bar{a}\bar{b}}, 
\end{equation} 
or 
\[
\dot{W} = \frac{1}{2} Q_{\bar{a}\bar{b}} \dot{\E}^{\bar{a}\bar{b}}
- \frac{1}{2} \frac{d}{dt} \Bigl(  Q_{\bar{a}\bar{b}}
   \E^{\bar{a}\bar{b}} \Bigr). 
\]
The total derivative corresponds to a change of a state function, and
we shall ignore such (reversible) changes in this analysis. Our final
expression for the rate of tidal work shall be 
\begin{equation} 
\dot{W} = \frac{1}{2} Q_{\bar{a}\bar{b}}\dot{\E}^{\bar{a}\bar{b}}. 
\label{eq:Wdot_def} 
\end{equation} 

If we insert Eq.~(\ref{eq:Q_inertial}) in this we find that the
$\Omega_H^2 C_{\bar{a}\bar{b}}$ and $\E_{\bar{a}\bar{b}}$ terms
contribute total derivatives, which we continue to ignore. The only
nontrivial contributions come from the terms proportional to $\tau$;
we get  
\begin{equation} 
\dot{W} = \frac{a^5 \tau}{4G}\, \dot{\E}^{\bar{a}\bar{b}} 
\bigl( \dot{\E}_{\bar{a}\bar{b}} + \Delta \dot{\E}_{\bar{a}\bar{b}}
\bigr). 
\label{eq:tidal_work} 
\end{equation} 
We recall that 
\begin{equation} 
\Delta \dot{\E}_{\bar{a}\bar{b}} = 
2\Omega_H \epsilon_{\bar{c}\bar{d}(\bar{a}} 
  \E^{\bar{c}}_{\ \bar{b})} s^{\bar{d}}
\label{eq:Delta_Edot} 
\end{equation} 
and 
\begin{equation} 
\tau = \frac{19}{2} \frac{a\nu}{GM}.  
\end{equation}  
The equation reveals that there is no net tidal work in the absence of 
viscosity. The dephasing between $Q _{\bar{a}\bar{b}}$ and
$\E_{\bar{a}\bar{b}}$ created by the viscous delay is therefore an
essential piece of the physics. In the absence of viscosity, the fluid 
displacement is always in phase with the applied force, and this leads
to a reversible transfer of gravitational energy between the body and
the tidal field; the work done necessarily averages to zero.  

For an external object of mass $M'$ on a circular orbit of radius $b$
and angular velocity $\Omega$, the rate of tidal work is 
\begin{equation} 
\dot{W} = \frac{a^5\tau}{18 G} \frac{(GM')^2}{b^6} \Omega (\Omega 
- \Omega_H). 
\end{equation} 
There is no work done if the external object is corotating with the
body. 

\subsection{Tidal torque} 

The total torque exerted by the tidal forces is 
\[
\tau_{\bar{a}} = \int \epsilon_{\bar{a}\bar{b}\bar{c}} x^{\bar{b}} 
\bigl(-\rho \E^{\bar{c}}_{\ \bar{d}} x^{\bar d} \bigr)\, d^3\bar{x}, 
\]
and this is equal to $\dot{J}_{\bar{a}}$, the rate of change of the
body's angular-momentum vector. Evaluation of the integral gives 
\begin{equation} 
\dot{J}_{\bar{a}} = -\epsilon_{\bar{a}\bar{b}\bar{c}}
Q^{\bar{b}}_{\ \bar{p}} \E^{\bar{p}\bar{c}}.
\end{equation} 
The angular-momentum vector can be decomposed as $J^{\bar{a}} 
= J s^{\bar{a}}$, in terms of a magnitude $J$ and a direction
$s^{\bar{a}}$. The change in angular momentum is then expressed as 
$\dot{J}^{\bar{a}} = \dot{J} s^{\bar{a}}+ J \dot{s}^{\bar{a}}$. The
first term describes the change in the body's rate of rotation, while
the second describes precessional effects. We are interested here in
the rate of change of $J$, and therefore select the projection of 
$\dot{J}^{\bar{a}}$ in the direction of $s^{\bar{a}}$. This is  
\begin{equation} 
\dot{J} = -\epsilon_{\bar{a}\bar{b}\bar{c}}
Q^{\bar{a}}_{\ \bar{p}} \E^{\bar{p}\bar{b}} s^{\bar{c}}.    
\label{eq:Jdot_def} 
\end{equation} 

When we insert Eq.~(\ref{eq:Q_inertial}) in this we find once more
that the $\Omega_H^2 C_{\bar{a}\bar{b}}$ and $\E_{\bar{a}\bar{b}}$
terms make no contribution. We obtain  
\begin{equation} 
\dot{J} = -\frac{a^5 \tau}{2G} 
\epsilon^{\bar{a}}_{\ \bar{p}\bar{q}} \E^{\bar{p}\bar{b}} s^{\bar{q}} 
\bigl( \dot{\E}_{\bar{a}\bar{b}} + \Delta \dot{\E}_{\bar{a}\bar{b}}
\bigr). 
\label{eq:tidal_torque1} 
\end{equation} 
We relate this to the rate of change of the body's rotational energy
$E_{\rm rot} = \frac{1}{2} I \Omega_H^2$, where $I$ is the body's
moment of inertia. With $J = I \Omega_H$ we find that 
$\dot{E}_{\rm rot} = \Omega_H \dot{J}$. Taking into account
Eq.~(\ref{eq:Delta_Edot}), Eq.~(\ref{eq:tidal_torque1}) becomes 
\begin{equation} 
\dot{E}_{\rm rot}= \Omega_H \dot{J} = 
-\frac{a^5 \tau}{4G} \Delta\dot{\E}^{\bar{a}\bar{b}}
\bigl( \dot{\E}_{\bar{a}\bar{b}}+ \Delta \dot{\E}_{\bar{a}\bar{b}}
\bigr). 
\label{eq:tidal_torque2} 
\end{equation}   
Again we find that the effect vanishes in the absence of viscosity: It
is the misalignment between the directions of the tidal forces and
tidal bulge that allows the forces to exert a torque on the body.  

For an external object of mass $M'$ on a circular orbit of radius $b$
and angular velocity $\Omega$, the rate of change of the body's
rotational energy is 
\begin{equation} 
\dot{E}_{\rm rot} = \Omega_H \dot{J} 
= \frac{a^5\tau}{18 G} \frac{(GM')^2}{b^6} 
  \Omega_H (\Omega - \Omega_H). 
\end{equation} 
There is no change if the external object is corotating with the
body. 

\subsection{Heat dissipation}  

The rate at which heat is dissipated within the fluid is given by
(see, for example, Sec.~16 of Ref.~\cite{landau-lifshitz:b6}) 
\begin{equation} 
\dot{Q} = \frac{1}{2} \rho \nu \int \bigl( 
\partial_{\bar{a}} v_{\bar{b}} + \partial_{\bar{b}} v_{\bar{a}} \bigr)  
\bigl( \partial^{\bar{a}} v^{\bar{b}} + \partial^{\bar{b}} v^{\bar{a}}
\bigr)\, d^3\bar{x}. 
\end{equation} 
The velocity field in the norotating frame is obtained by transforming 
Eq.~(\ref{eq:v_field}) using the rules spelled out in Sec.~III E. We
have 
\begin{eqnarray} 
v^{\bar{a}} &=& \Omega_H \epsilon^{\bar{a}}_{\ \bar{b}\bar{c}}
s^{\bar{b}} x^{\bar{c}} + \frac{a}{4GM} \Bigl[ (5\bar{r}^2 - 8 a^2) 
\bigl( \dot{\E}^{\bar{a}}_{\ \bar{b}}
+ \Delta \dot{\E}^{\bar{a}}_{\ \bar{b}} \bigr) x^{\bar{b}} 
\nonumber \\ & & \mbox{}
- 2 \bigl( \dot{\E}_{\bar{b}\bar{c}} 
+ \Delta \dot{\E}_{\bar{b}\bar{c}} \bigr) x^{\bar{a}}
  x^{\bar{b}} x^{\bar{c}} \Bigr], 
\end{eqnarray}
and this can be inserted within our expression for $\dot{Q}$.  

A straightforward computation returns   
\begin{equation} 
\dot{Q} = \frac{a^5 \tau}{4G} \bigl( 
\dot{\E}^{\bar{a}\bar{b}} + \Delta\dot{\E}^{\bar{a}\bar{b}} \bigr) 
\bigl( \dot{\E}_{\bar{a}\bar{b}}+ \Delta \dot{\E}_{\bar{a}\bar{b}}
\bigr). 
\label{eq:tidal_heat} 
\end{equation} 
Notice that this is positive-definite, as it should be. 

For an external object of mass $M'$ on a circular orbit of radius $b$
and angular velocity $\Omega$, the rate at which heat is dissipated
within the fluid is 
\begin{equation} 
\dot{Q} 
= \frac{a^5\tau}{18 G} \frac{(GM')^2}{b^6} 
   (\Omega - \Omega_H)^2. 
\end{equation} 

\subsection{Energy balance} 

The work done by the tidal forces is used partially to increase the
body's rotational energy, and is partially dissipated into heat by the
fluid's viscosity. Energy balance requires that 
\begin{equation} 
\dot{W} = \dot{Q} + \Omega_H \dot{J}, 
\label{eq:en_balance} 
\end{equation} 
and we see from Eqs.~(\ref{eq:tidal_work}), (\ref{eq:tidal_torque2}),
and (\ref{eq:tidal_heat}) that the equation is indeed satisfied. 

\subsection{Generalization to arbitrary internal structure and
  correspondence with black-hole physics} 

The results derived previously apply specifically to an incompressible 
fluid. It is easy, however, to generalize them so that they apply
to a fluid body with an arbitrary internal structure (generated by an 
arbitrary equation of state). The trick is to preserve the main
structure of each equation, including the scaling with each
dimensionful quantity, but to insert various dimensionless
coefficients where appropriate. These coefficients are called
``Love numbers,'' and their role is precisely to incorporate the
details of the body's internal structure. This strategy is viable so
long as one is interested only in the structure and scalings of the
main equations, and one is satisfied not to know the exact value of
each Love number. A detailed computation of the Love numbers for a
selected equation of state would require the hard work of repeating
the analysis of fluid perturbations for this equation of state.  

The first result that can be generalized in this way is
Eq.~(\ref{eq:Qrot}), which relates $Q_{jk}$, the quadrupole moment of
the body's mass distribution, to $\E_{jk}$, the quadrupole moment of
the tidal field. The generalization appears in
Eq.~(\ref{eq:Q_vs_E1}); it involves a rotational Love number $n_2$, a
tidal Love number $k_2$, and a viscous delay $\tau$ that is known to
scale as $a\nu /(GM)$ with the body radius $a$, its mass $M$, and its
averaged kinetic viscosity $\nu$. For an incompressible fluid with a
uniform viscosity, $n_2 = k_2 = \frac{3}{4}$ and $\tau = \frac{19}{2}
a\nu/(GM)$. [In the generalization I chose to promote $\tau$ to a
dimensionful ``Love quantity.'' An alternative would have been to
define a new Love number $v_2$ and to generalize $\tau$ as $v_2
a\nu/(GM)$. The former strategy is more useful for the purpose of
establishing a correspondence with the black-hole results.] Similarly,
the generalization of Eq.~(\ref{eq:Q_inertial})  appears in
Eq.~(\ref{eq:Q_vs_E2}).  

The expressions obtained for $\dot{W}$ and $\dot{J}$ --- see
Eqs.~(\ref{eq:tidal_work}) and (\ref{eq:tidal_torque2}), respectively
--- follow directly from the basic definitions of
Eqs.~(\ref{eq:Wdot_def}), (\ref{eq:Jdot_def}) and the relation of
Eq.~(\ref{eq:Q_inertial}) between $Q_{\bar{a}\bar{b}}$ and 
$\E_{\bar{a}\bar{b}}$. The definitions continue to apply even after
Eq.~(\ref{eq:Q_inertial}) is generalized to Eq.~(\ref{eq:Q_vs_E2}),
and we arrive at 
\begin{equation} 
\dot{W} = \frac{(k_2 \tau) a^5}{3G}\, \dot{\E}^{\bar{a}\bar{b}}   
\bigl( \dot{\E}_{\bar{a}\bar{b}} + \Delta \dot{\E}_{\bar{a}\bar{b}}
\bigr) 
\label{eq:Wdot_gen} 
\end{equation} 
and 
\begin{equation}  
\Omega_H \dot{J} = -\frac{(k_2 \tau)a^5}{3G}
\Delta\dot{\E}^{\bar{a}\bar{b}}    
\bigl( \dot{\E}_{\bar{a}\bar{b}}
+ \Delta \dot{\E}_{\bar{a}\bar{b}} \bigr).  
\label{eq:Jdot_gen} 
\end{equation}  
These are Eqs.~(\ref{eq:WdotN}) and (\ref{eq:JdotN}), respectively (in
which the overbar is removed from the indices to simplify the
notation, and $G$ is set equal to unity). We can then invoke the
energy-balance equation (\ref{eq:en_balance}) to motivate the
generalized form of Eq.~(\ref{eq:tidal_heat}),  
\begin{equation} 
\dot{Q} = \frac{(k_2 \tau) a^5}{3G} \bigl( 
\dot{\E}^{\bar{a}\bar{b}} + \Delta\dot{\E}^{\bar{a}\bar{b}} \bigr) 
\bigl( \dot{\E}_{\bar{a}\bar{b}}+ \Delta \dot{\E}_{\bar{a}\bar{b}}
\bigr).  
\label{eq:Qdot_gen} 
\end{equation} 

Equations (\ref{eq:Wdot_gen}), (\ref{eq:Jdot_gen}), and
(\ref{eq:Qdot_gen}) can be compared with the corresponding black-hole 
results displayed in Sec.~I. We associate $\dot{W}$ with $\dot{M}$,
$\dot{J}$ with $\dot{J}$, and $\dot{Q}$ with $(\kappa/8\pi) \dot{A}$,
where $\kappa$ is the black hole's surface gravity and $A$ its surface 
area. With this association the energy-balance equation of
Eq.~(\ref{eq:en_balance}) becomes $\dot{M} = 
(\kappa/8\pi) \dot{A} + \Omega_H \dot{J}$, the first law of black-hole 
mechanics. The details of this comparison were already described in
Sec.~I.  

\begin{acknowledgments} 
This work was supported by the Natural Sciences and Engineering
Research Council of Canada. I thank Eanna Flanagan and Tanja Hinderer
for a useful discussion, and Simon Comeau for inspiring me to complete
this work. I am also grateful to the staff of the Canadian Institute
for Theoretical Astrophysics for their kind hospitality during the
time of my research leave from the University of Guelph; a large of
portion of this work was completed during this time.      
\end{acknowledgments} 

\bibliography{../bib/master} 
\end{document}